\documentclass[twocolumn,secnumarabic,amssymb, nobibnotes, aps, prd, showkeys]{revtex4-2}
\usepackage{amssymb}
\usepackage{graphics}
\usepackage{color}
\usepackage{epsfig}
\usepackage{verbatim}
\usepackage{bm}
\usepackage{multirow}

\begin{document}

\title{Impact of terahertz short pulses on the oxygen defect state in TiO$_{2-x}$}

\author{Paola Di Pietro$^{1}$, Johannes Schmidt$^{1}$, Nidhi Adhlakha$^{1}$, Sandeep Kumar Chaluvadi$^{2}$, Federico Mazzola$^{2,3}$, Veronica Stopponi$^{2}$, Luca Tomarchio$^{4}$, Pasquale Orgiani$^{2}$, Stefano Lupi$^{2}$ and Andrea Perucchi$^{1}$}

\affiliation{$^1$Elettra - Sincrotrone Trieste S.C.p.A, S.S. 14 km163.5 in AREA Science Park, 34012 Trieste, Italy}
\affiliation{$^2$CNR-IOM – Istituto Officina dei Materiali, 34149 Trieste, Italy}
\affiliation{$^3$Department of Molecular Sciences and Nanosystems, Ca’ Foscari University of Venice, I-30172 Venice, Italy}
\affiliation{$^4$Dipartimento di Fisica, Universit\`a di Roma Sapienza,  P.le Aldo Moro 2, I-00185 Roma, Italy}

\date{\today}

\begin{abstract}

Oxygen deficient titanium dioxide (TiO$_{2-x}$) is a very attractive material for several applications ranging from photocatalysis to resistive switching. Oxygen vacancies turn insulating anatase titanium dioxide into a polaronic conductor, while creating a defect state band below the ultraviolet semiconducting gap. Here we employ a combination of broadband infrared (IR) reflectivity and THz-pump/IR-probe measurements to investigate the relationship between localized defect states and delocalized conducting polaronic states. We show that the THz pump allows to convert deeply localized electrons into metastable polarons with a lifetime in the ns range. These long-lived metastable states may find application in novel opto-electronic applications exploiting the interplay of dc resistivity, with terahertz and infrared signals.

\end{abstract}


%\pacs{}

\maketitle

\section{Introduction}
\label{introduction}

Titanium dioxide TiO$_2$ is a largely available, naturally occurring oxide which is being used for many applications since hundreds of years. It is chemically inert and semiconducting, and exhibits photocatalytic activity in the presence of light with energy equal to or higher than its electronic band-gap \cite{hashimoto, fujishima}. Its photocatalytic activity allows for thin coatings exhibiting self-cleaning and disinfecting properties under exposure to ultraviolet radiation. Due to its large (3.2 eV) band-gap, TiO$_2$ is also used as a white pigment for the preparation of enamels, cosmetics, and sunscreens. Exciton dynamics in stoichiometric TiO$_2$ has been extensively studied with above gap UV pump-probe technique \cite{baldini1, baldini2, baldini3, baldini4, baldini5}. Nonetheless, for some photocatalytic applications the ultraviolet band-gap represents a limiting factor, because of the limited portion of the solar spectrum (5 \%) exceeding the interband transition threshold \cite{oregan}.

The introduction of oxygen vacancies (V$_{\text{O}}$s) through annealing procedures brings in very important changes to the electrodynamic properties of TiO$_2$. The onset of a sizeable Drude conducting term is accompanied by the appearance of infrared absorption bands \cite{orgiani20}. While those terms originate all from the oxygen defects, it is far from obvious whether they belong to the same type of V$_{\text{O}}$. Angle resolved photoemission together with DFT (Density Functional Theory) calculations suggest on the contrary the coexistence of distinct localized and delocalized states, with different spatial location and diffusion kinetics \cite{bigi20, selloni}. 

Oxygen vacancies in TiO$_{2-x}$ are being studied extensively due to their ability to enhance the photocatalytic activity of TiO$_{2-x}$, which is important for various applications such as photocatalytic degradation of pollutants and water splitting for hydrogen production. V$_{\text{O}}$s also play a prominent role in the engineering of resistive switching properties, thereby allowing the use of TiO$_{2-x}$ as a memristor \cite{xiao22,lee11,kousar21,szot11, kope11} for in-memory computing schemes.

We employ here a sub-gap terahertz (THz) excitation to achieve control on the oxygen defect state, thereby modulating the infrared response associated to the photo-excitation of V$_{\text{O}}$s. Our work highlights the presence of very different relaxation dynamics. A fast response (sub-ps) closely follows the THz excitation, after which a very long (ns) relaxation dynamics takes place, associated to the formation of a metastable state. The dynamics unveiled by our THz-pump/IR-probe experiment allows to interpret this transient state in terms of a conversion from deep localized defect states to delocalized large polarons, and back \cite{calvani01, emin93}. This finding has important consequences towards the exploitation of TiO$_{2-x}$ for solar energy conversion, as well as in novel opto-electronic applications, for instance designing memristors realizing high-bandwidth neuromorphic vision \cite{shan21}.

\begin{figure}
\includegraphics [width=8.5cm]{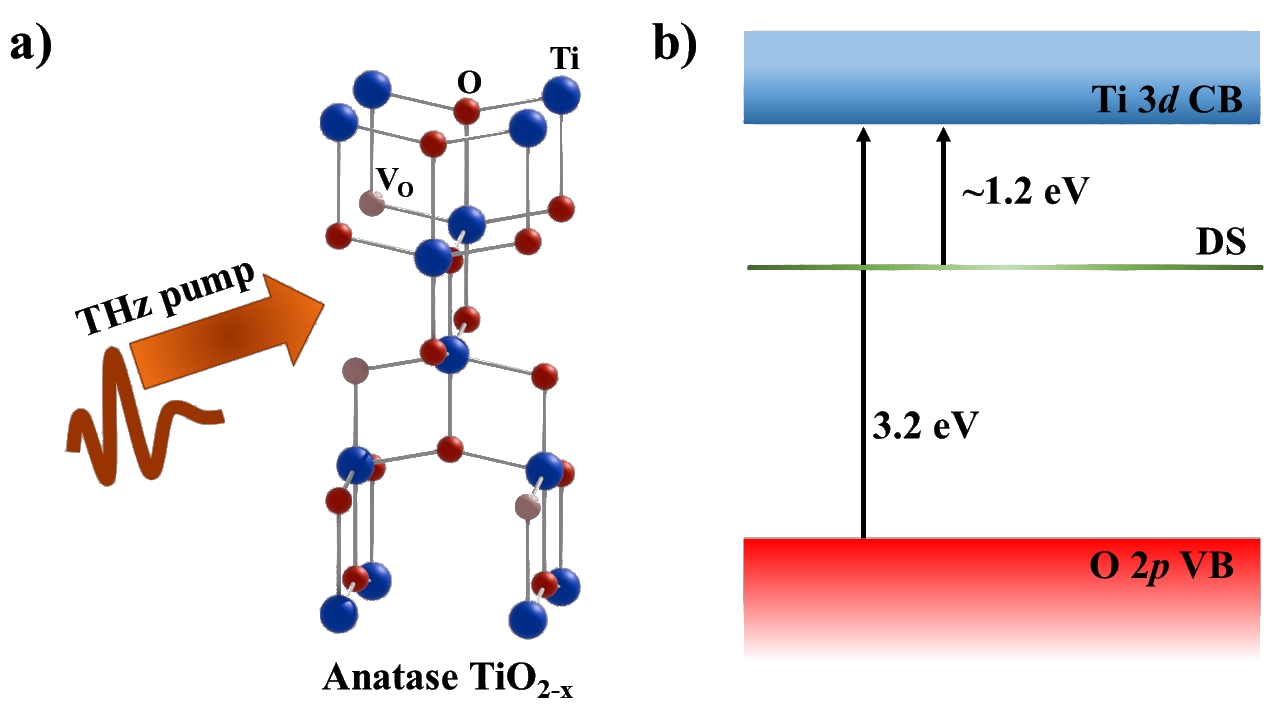}  
\caption{a) Crystal structure of anatase TiO$_{2-x}$. b) Schematic distribution of the electronic bands in oxygen deficient TiO$_2$, where DS indicates the defect state.}
\label{Fig1}
\end{figure}

\section{Steady state characterization}
\label{FTIR}

\begin{figure*}
\leavevmode
\includegraphics [width=10cm]{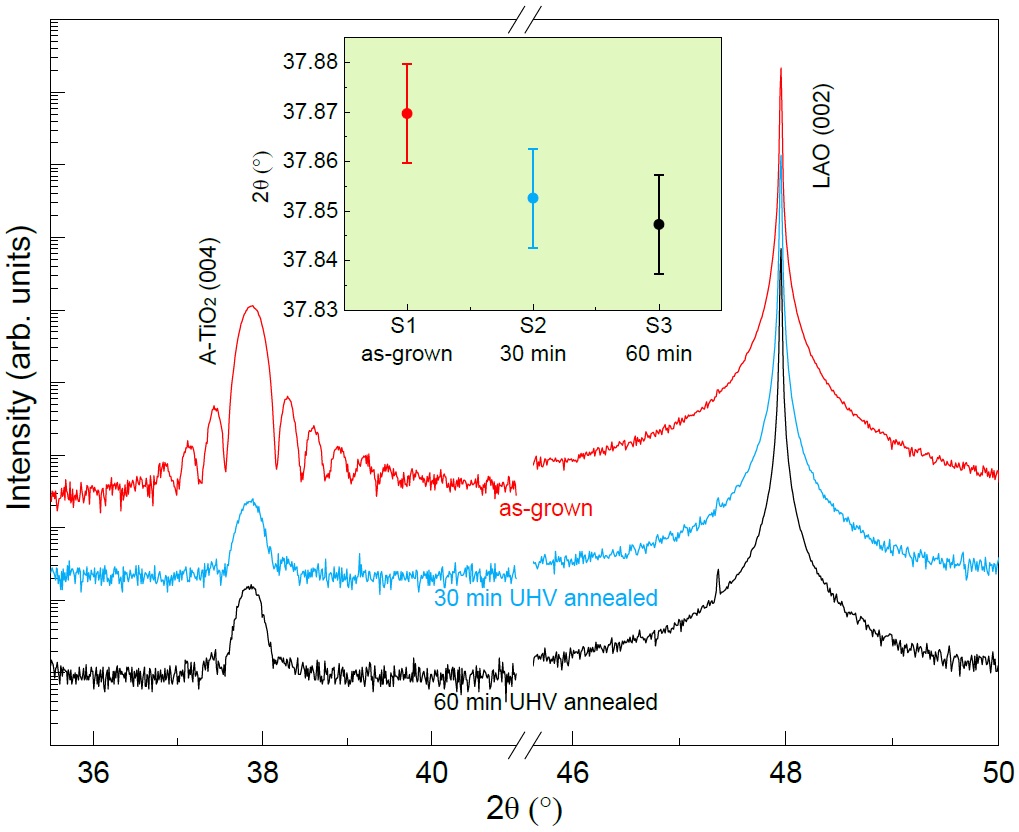}  
\caption{$\theta-2\theta$ symmetrical XRD scans of anatase TiO$_{2-x}$ thin films deposited on LaAlO$_3$ (substrates) under different post-annealing conditions, respectively. Inset shows the (004) A-TiO$_2$ peak position of three different films.}
\label{figXRD}
\end{figure*}

Very high crystalline quality of anatase TiO$_{2-x}$ thin films of 40 nm thick were deposited on double-side polished LaAlO$_3$ (001) substrates by pulsed laser deposition (PLD) technique (see Section \ref{mm}). In this study, three samples -- namely as-grown (S1), 30 minutes (S2) and 60 minutes (S3) -- UHV post-annealed films were chosen \cite{orgiani20, Knez20}. The crystalline quality of the films was measured by a x-ray diffractometer as shown in Fig.  \ref{figXRD}. The $\theta-2\theta$ XRD of the as-grown film shows Laue oscillations indicating very high crystalline quality of the film. For the UHV post-annealing samples, two features were observed: i) Laue oscillations tend to disappear owing to increase in structural disorder, and ii) red shift in the $2\theta$ peak position of the (004) A-TiO$_2$ (inset of Fig. \ref{figXRD}) corresponding to the expansion of out-of-plane lattice constant "c" \textit{i.e.}, increase in oxygen vacancies in the film \cite{orgiani20}. 

%\section{Optical Spectroscopy}
\begin{figure*}
\leavevmode
\includegraphics [width=15cm]{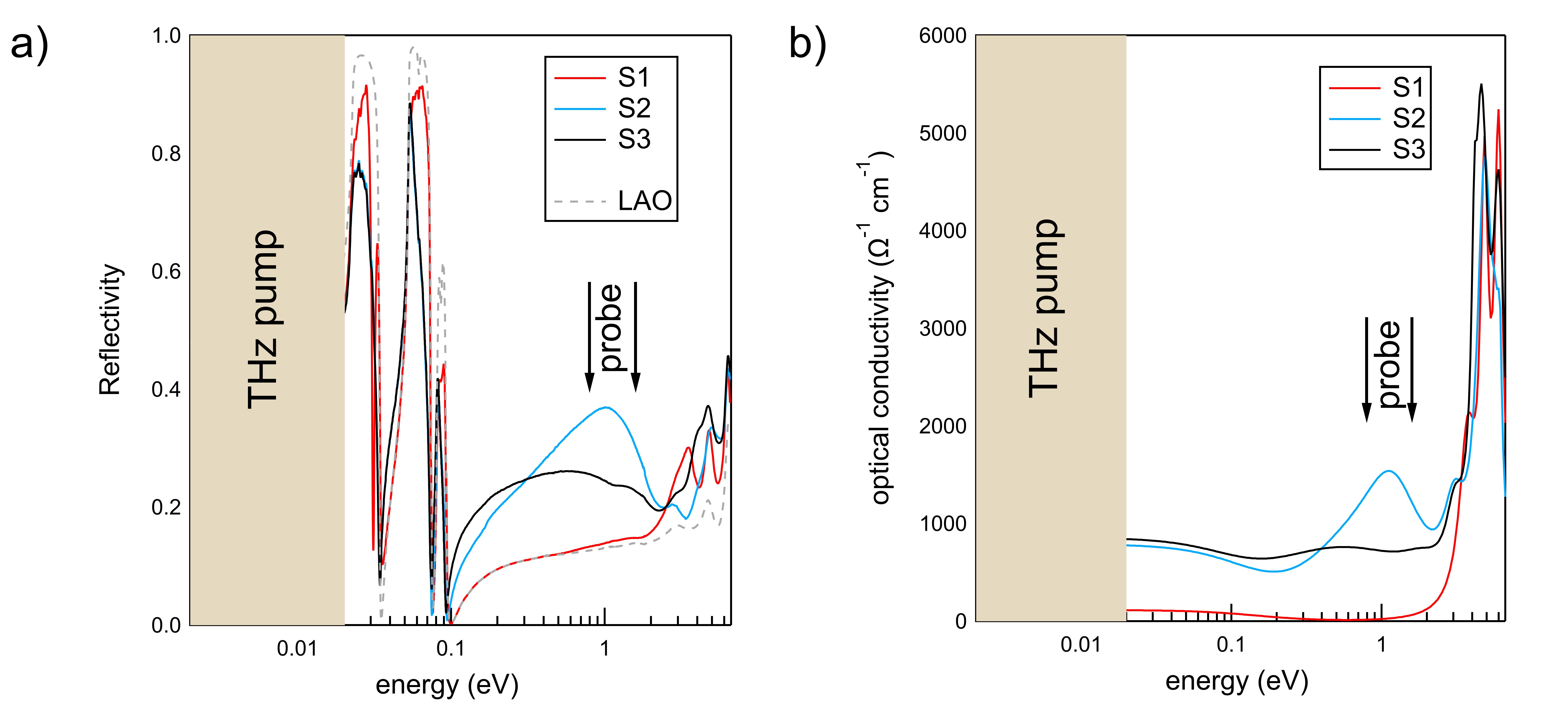}  
\caption{a) Optical reflectivity measurements performed on the S1, S2 and S3 TiO$_{2-x}$ thin films, deposited on LaAlO$_3$ (001) substrate. b) Real part of the optical conductivity, as extracted from a Drude-Lorentz modelling of the data in a).}
\label{fig2}
\end{figure*}

Broadband optical reflectivity measurements have been performed from the THz (3 THz $\sim$ 12 meV) to the ultraviolet ($\sim$ 6.5 eV) spectral range on S1, S2 and S3 samples.
In the far-infrared range, \textit{i.e.} below $\sim$ 0.1 eV, the reflectivity is dominated by the infrared phonon modes (Reststrahlen bands) of the LaAlO$_3$ substrate, which are only partially screened by the thin TiO$_{2-x}$ film (Fig. \ref{fig2}a). The more conducting is the film, the lower will be the reflectivity of the Reststrahlen band components. At higher frequencies, the enhancement of the sample reflectivity with respect to that of the substrate reflects the onset of absorption bands. In particular, a clear feature peaked at about 1 eV is present in the reflectivity of sample S2. At even higher frequencies, the reflectivity exhibits other peaks, associated to the electronic structure of both the samples and the substrate.

\begin{figure}
\includegraphics [width=8cm]{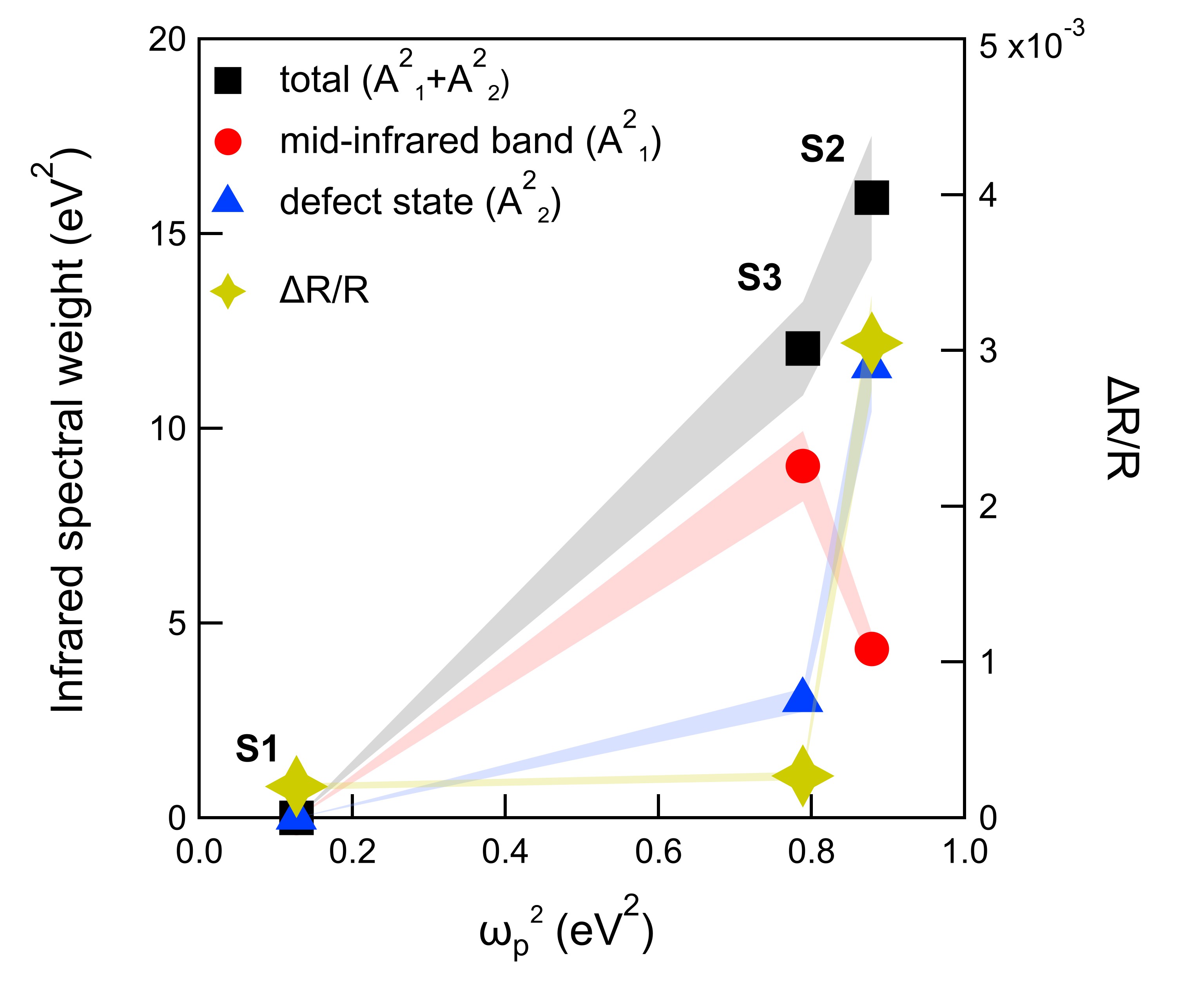}  
\caption{Spectral weight of the infrared components ($A_1^2$, $A_2^2$, and their sum $A_1^2+A_2^2$) from the Drude-Lorentz fitting, plotted as a function of the Drude spectral weight $\omega_p^2$. $\Delta R/R$ integrated within the first 30 ps is reported on the right scale.}
\label{figSW}
\end{figure}

The reflectivity data are fitted within a Drude-Lorentz model \cite{dressel} describing the complex dielectric function $\tilde{\epsilon}(\omega)$ in the form:
\begin{equation}
\tilde{\epsilon}(\omega)=1-\frac{\omega_p^2}{\omega^2+i\omega/\tau}+\Sigma_{i=1}^6\frac{A_i^2}{(\omega_{0,i}^2-\omega^2)-i\omega/\tau_i}
\label{drudelorentz}
\end{equation}
Here $\omega_p$ is the Drude plasma frequency ($\omega_p^2$ is the Drude spectral weight), while $A_i$ represent the oscillator strengths of the minimal set of six lorentzian components necessary to reproduce our data, two for intra-gap bands (a mid-infrared and a near-infrared) and four for ultraviolet inter-gap bands. The fitting formula takes into account the reflection from both the thin film and the substrate within the Fresnel formalism, as already done in Ref. \onlinecite{orgiani20}. For all samples the optical conductivity above 3.2 eV suddenly increases as a consequence of the onset of the electronic band gap. On the low frequency side, one can clearly distinguish the presence of an overdamped Drude term in the annealed samples (S2 and S3). The infrared spectral weight is taken into account here by a combination of two lorentzian components, at about 0.5 ($\omega_{0,1}$) and 1.2 eV ($\omega_{0,2}$). The sum of these two components ($A_1^2+A_2^2$) is reported in Fig. \ref{figSW}, as a function of the Drude spectral weight $\omega_p^2$.

%\begin{table} [ht!]
% \centering
 % \begin{tabular}{|c |c || c | c |c |c |}
% \hline
% & & A & B & C & D \\
 %\hline
%\hline
 %Drude & $\omega_p$ (eV)  & 0.94    & 0.89   &  0.68  &   0.36     \\
%\hline
%\hline
 %\multirow{2}{*}{Mid-IR}  & $\omega_{0,1}$  (eV)&   0.59  & 0.58   & 0.66   &  0   \\
%\cline{2-6}
    %             & $A_1$ (eV) & 2.08  &  3  &  2.44  &   0     \\
%%\hline
%\hline
%\multirow{2}{*}{DS}  & $\omega_{0,2}$  (eV)& 1.18    &  1.86  & 0   &   0    \\
%\cline{2-6}
    %            & $A_2$ (eV) & 3.40    &  1.74  & 0   &   0       \\
%\hline
 % \end{tabular}
% \caption{Fitting parameters of the Drude, the mid-infrared and the defect state bands. Here, $\omega_p$ refers to the plasma frequency, $\omega_{0,(1,2)}$ to the central frequency of the mid-infrared and the defect state band respectively and $A_{(1,2)}$ to their oscillator strengths, as defined in equation(\ref{drudelorentz}).}
 %\label{para}
%\end{table} 

The infrared spectral weight monotonically grows with the increase of the Drude term. By looking more in detail at the fitting components one notices that the mid-infrared band intensity (red dots) reaches its maximum for sample S3, whereas in sample S2 the component providing the largest contribution to the infrared spectral weight is that at 1.2 eV (blue triangles). It is tempting to interpret this behavior of the optical conductivity within the traditional dichotomy between deep and shallow donors, which is often discussed in TiO$_{2-x}$ literature \cite{mattioli10}.

The coexistence of a Drude term together with a mid-infrared band is an ubiquitous phenomenon in transition metal oxides. In a polar lattice as that of anatase TiO$_2$ the mid-infrared absorption is usually explained with polaron formation. Excess electrons delocalized over several Ti atoms interact with O$^{2-}$ ions, thereby inducing a distortion of the lattice, which can be seen as a large polaron \cite{emin93,calvani01}

In oxygen deficient TiO$_2$, the infrared spectral signature of the presence of oxygen vacancies is represented by a near-infrared feature which accounts for the structural lattice distortion induced by the vacancy formation, attracting a deep localized electron (usually referred as defect state, DS). The redistribution of spectral weight between those two bands in S2 and S3 well agrees with the fact that the two annealed samples contain similar amount of V$_{\text{O}}$s, as it has been measured by XRD (see inset in Fig. \ref{figXRD}).

To corroborate our attribution and in order to better understand the behaviour of the electrons surrounding the oxygen vacancy, we address time-resolved THz-pump/IR-probe spectroscopy. This technique allows exciting electrons localized at the oxygen vacancy by probing the variation of the sample reflectivity at frequencies corresponding to the mid-infrared and defect state excitations.

\begin{figure}
\includegraphics [width=8cm]{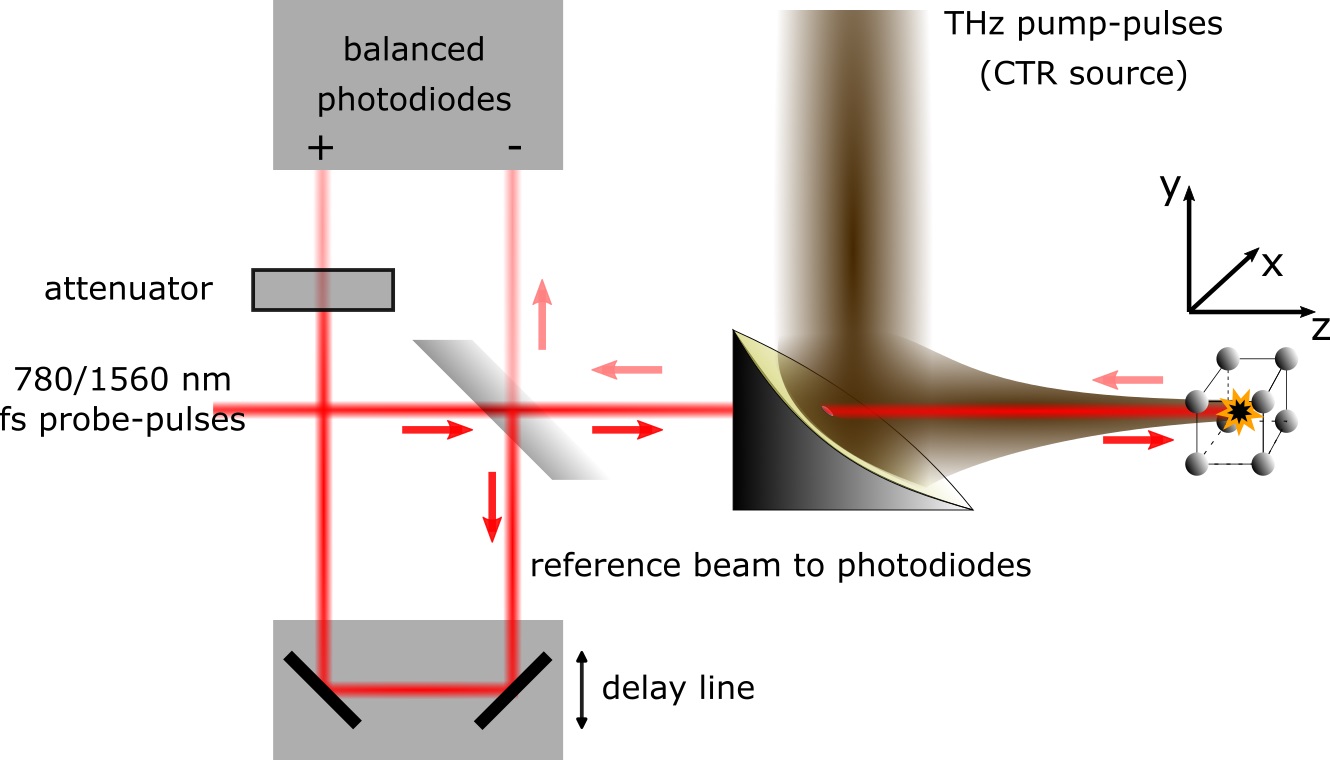}  
\caption{Sketch of the experimental set-up available at TeraFERMI for THz-pump/IR-probe measurements.}
\label{setup}
\end{figure}

\section{Time-Resolved spectroscopy}

\begin{figure*}
\leavevmode
\includegraphics [width=15cm]{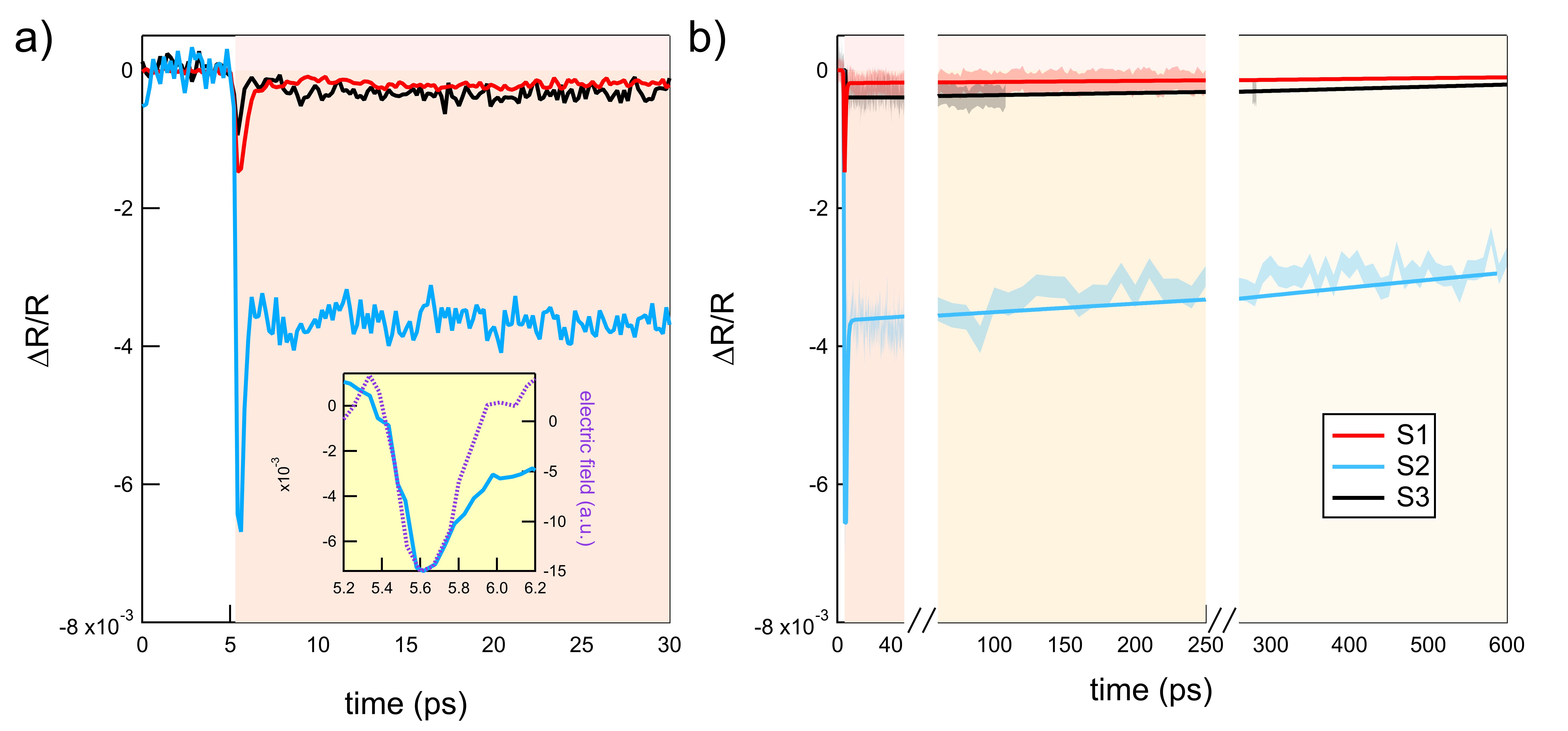}  
\caption{Transient reflectivity $\Delta R/R$ measured at 780 nm, with a THz pump field of 5 MV/cm. Panel a) shows data in the first 25 ps after the pump arrival, while panel b) display results with delays up to 600 ps. The inset in panel a) shows a comparison of the THz pump field (dashed purple line) and $\Delta R/R$ of sample S2 over a short time-range. In panel b) continuous lines are fits to the experimental data which are shown here in shaded colors.}
\label{figpptest}
\end{figure*}

THz-pump/IR-probe measurements have been performed at the TeraFERMI beamline \cite{TeraFERMI, perucchi13} of the FERMI facility. TeraFERMI provides broadband (0.3 to 5 THz), short ($\sim$ 10$^2$ fs) and very intense (electric field up to 5 MV/cm) single-cycle pulses to be used as pump beam. The IR probe pulses are provided by an optically synchronized fiber laser within 66 fs jitter \cite{Roussel} with the terahertz radiation. The laser can be employed both at its 1560 nm (0.8 eV) fundamental wavelength or at its second harmonic 780 nm (1.6 eV). A sketch of the set-up is shown in Fig. \ref{setup}.
The transient reflectivities of the S1, S2 and S3 samples, measured with the 780 nm probe, are shown in Fig. \ref{figpptest}a in a time range of 30 ps. The highest response is given by the sample S2, where the static reflectivity at 1.6 eV is the highest (see Fig. \ref{fig2}), while the S1 and S3 samples provide comparable responses. We compare in Fig. \ref{figSW} the area of the $\Delta R/R$ transient response (integrated over the time interval 0-30 ps), with the spectral weight of the various components. It clearly turns out that $\Delta R/R$ directly scales with the intensity of the defect state component, whereas all other fitting components follow distinctly different behaviours. This confirms that the 780 nm probe measurement explores the dynamics of the defect state after THz photo-excitation. Indeed, since the Drude spectral weight is also not scaling linearly with $\Delta R/R$, we can safely rule out a prominent contribution from free charge carriers to the pump-probe signal.

%Since both the 780 nm (1.6 eV) and 1560 nm (0.8 eV) wavelength fall well into the broad near-infrared absorption band, present in the static reflectivity of the annealed samples (see Figure \ref{fig2}), the IR laser beam probes the variation of the transient reflectivity at the defect state after the THz pump excitation.

\begin{figure*}
\leavevmode
\includegraphics [width=15cm]{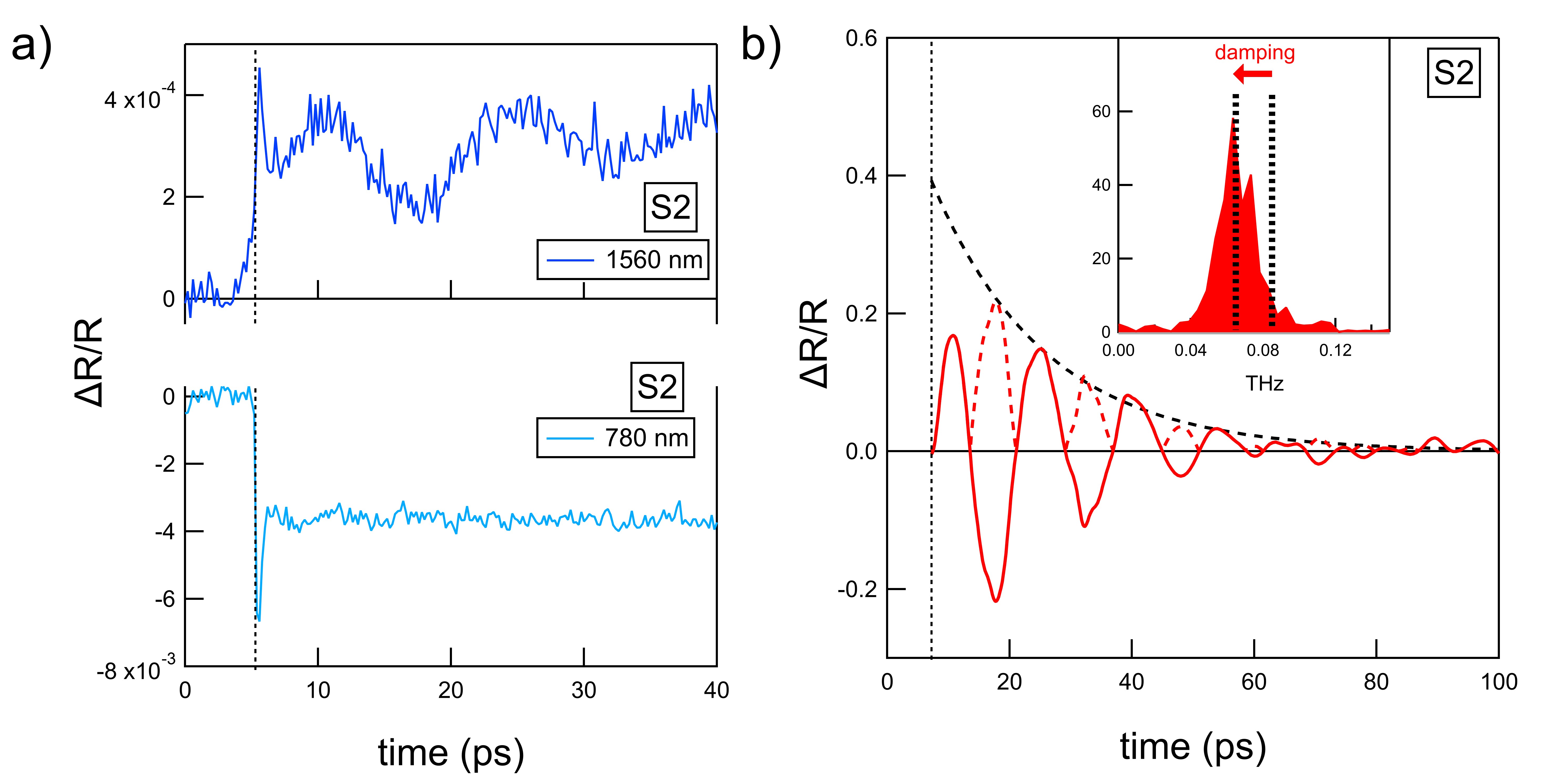}  
\caption{a) Transient reflectivity $\Delta R/R$ measured on sample S2 at 1560 nm, with a THz pump field of 1.25 MV/cm (top panel). $\Delta R/R$ measured at 780 nm ($E_{THz}=$5 MV/cm, see Section \ref{mm}) is also shown for comparison (bottom panel). b) Oscillations extracted from the curve reported in a) (top panel), as described in Section \ref{mm}. The dashed black line represents the exponential envelope of the damped oscillations. The inset shows the Fourier Transform of the oscillations affected by the damping constant, which causes a red shift of the resonance frequency from $f_0$ = 85 GHz to $f_{\text{D}}$ = 65 GHz (see main text).}
\label{A_1.5um}
\end{figure*}

%\textcolor{red}{Surprisingly, all the transient reflectivities are negative but that referred to sample C. This means that in A, B and D the reflectivity decreases under the THz pump excitation, while in C the reflectivity increases. This trend depends on the softening (hardening) of the DS band below the excitation. HERE WE HAVE TO DECIDE WHETHER TO KEEP INSIDE THE SAMPLE C, WITHOUT HAVING YET A CLEAR EXPLANATION FOR ITS DIFFERENT BEHAVIOUR OR TO REMOVE IT.}

The THz-pump/780 nm-probe data, shown in Fig. \ref{figpptest}, exhibit the presence of at least three different time scales. A first one is associated to the excitation from the THz pulse, while two very different timescales for relaxation are present.
The rise-time after photo-excitation is very sharp for all samples. A sigmoidal fit of this feature provides values $\leqslant 100$ fs, which corresponds to the rise time of the THz pump electric field, measured separetely by Electro-Optic Sampling (EOS). Within the same time scale is the pulse width of IR probe pulses, here 110 fs, which determines the time resolution of our experiment. One can safely assume that the electronic excitation process takes place in timescale shorter than 100 fs for all samples.
The relaxation processes occurring after photo-excitation are being fitted by a standard double exponential model. The short relaxation ($\tau_{short}$) corresponds to 200, 400 and 260 fs for samples S1, S2 and S3, respectively. It is worth mentioning that for all samples, the relaxation $\tau_{short}$ is longer than the decay of the (almost) symmetric single cycle THz pump pulse. This is better seen in the inset of Fig. \ref{figpptest}a, comparing the pump-probe signal of sample S2, with the THz electric field. Here, one may notice that while the rise of the $\Delta R/R$ signal matches the rise of the THz pulse, $\Delta R/R$ decays more slowly than the THz electric field.
After this first sharp decay, the $\Delta R/R$ becomes almost flat for all samples. If measured over a much longer ($\sim 10^2$ ps) time delay as shown in Fig. \ref{figpptest}b, one can distinguish a long relaxation ($\tau_{long}$) assuming values in the order of ns.

The transient reflectivity of sample S2 has also been measured by employing a 1560 nm probe, in order to investigate the dynamics at a wavelength falling in the spectral region between $A_1$ and $A_2$ bands. The response has a positive sign (Fig. \ref{A_1.5um}a), contrary to the signal probed at 780 nm. As in the case of the 780 nm measurements, the transient reflectivity includes different dynamical responses, which can be analyzed separately. A first very short and sharp rise is identifiable, after which a long step-like dynamics occurs, with time constant comparable with that found at 780 nm. Moreover, clear oscillations superimpose to the step-like response and are fully damped after $\sim$50 ps. 
%The Fourier transform of this temporal oscillating response is peaked at 63 GHz (see inset of Fig. \ref{A_1.5um}). 
We will come back to this feature in the next Section \ref{discussion}.

\section{Discussion}
\label{discussion}
\begin{figure}
\includegraphics [width=8.5cm]{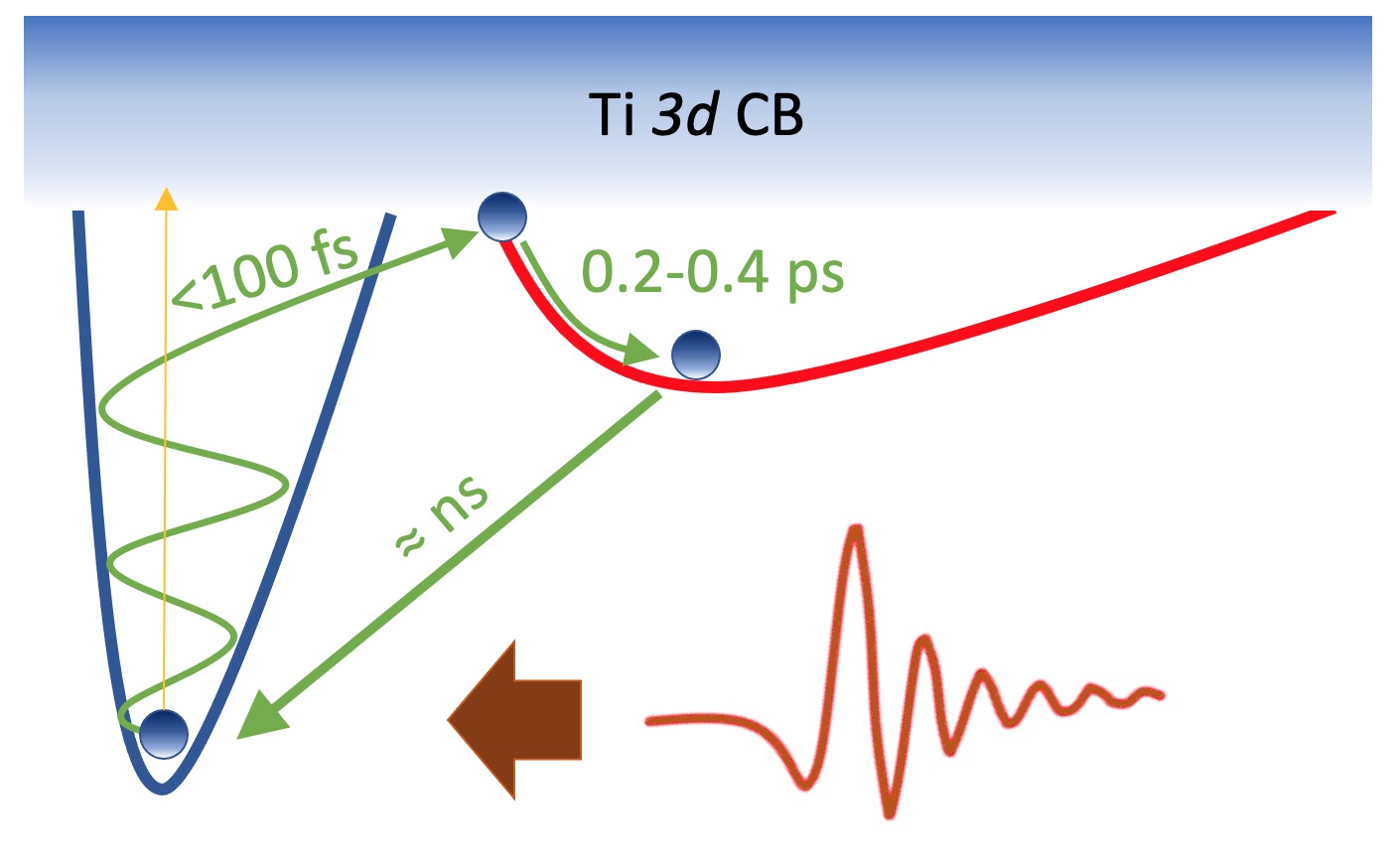}  
\caption{Sketch of the THz excitation of a deep localized vacancy. The electron localized in the vacancy acquires sufficient energy from the THz field, leading to its photoionization within few 10's fs. A large polaron is then formed within less than 1 ps. This represents a metastable state requiring $\sim 1$ ns before relaxing back to the strongly localized vacancy ground state.}% qui mettere eventualmente riferimento a fig.1
\label{polaron}
\end{figure}

The most intuitive interpretation of our experimental data is in terms of polaron excitation by a THz field, as described in the sketch of Fig. \ref{polaron}. The defect state can be seen as a small polaron localized at the oxygen vacancies. The pumping THz field interacts with both the electrons and the ions thereby providing a displacement strong enough as to climb the defect state potential well \cite{jerawa10}, and eventually lead to electrons escape from the oxygen vacancy site \cite{kornilovitch17}. The photoionized free charge carrier is now allowed to delocalize within the Ti$-3d$ conduction band. The underlying strongly polar lattice leads to large polaron formation, accompanied by a structural distortion extended over several Ti sites. Consistent with previous literature \cite{bretschneider18}, the formation of the polaron employs from 200 to 400 fs, which is the time interval corresponding to $\tau_{short}$. The large polaron represents a metastable state employing few ns ($\tau_{long}$) to relax back to the original defect state. This time-scale relates to the time needed for a scattering event to take place between the large polaron and an oxygen vacancy, thus allowing the relaxation to the deep defect state to occur. 
%The spatial separation between the negatively charged large polaron and the states provided by the positive V$_{\text{O}}$ therefore plays a role in determining the ns relaxation $\tau_{long}$.

%Interestingly $\tau_{long}$ is larger on the more deficient samples, and decreases monotonically with decreasing number of V$_{\text{O}}$s. This observation is apparently at odds with the interpretation of $\tau_{long}$ proposed above. Actually, as remarked in Section \ref{FTIR}, the increased concentration of V$_{\text{O}}$s is also expected to increase clustering, in the form of conducting filamentary chains. The formation of these self-ordered structure may decrease the probability of scattering events between the large polarons and V$_{\text{O}}$s, even if the total number of V$_{\text{O}}$s increases.

%qui inserire plot tau_long

\begin{figure}
\includegraphics [width=8cm]{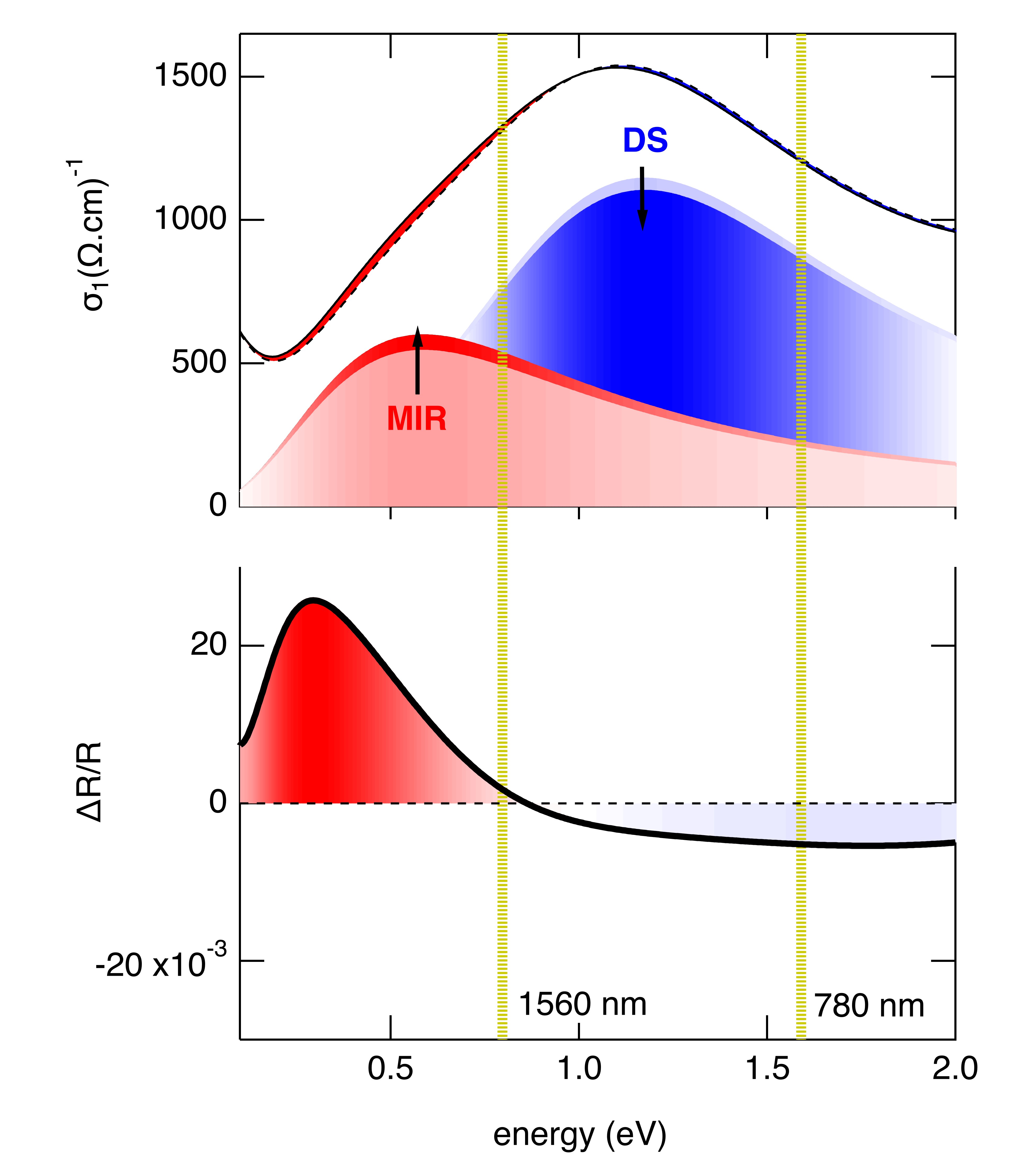}  
\caption{a) Variation (coloured) of the optical conductivity $\sigma_1$ of sample S2 from that extracted from the FTIR measurement to that simulated {\it after} a THz pump beam has reduced the spectral weight component $A_2$ related to the defect state (DS) excitation, while the component $A_1$ (mid-infrared, MIR) simultaneously increases in order to conserve the total spectral weight b) Corresponding calculated $\Delta R/R$. The yellow vertical lines indicate the 780 nm and 1560 nm.}
\label{sigma_pp}
\end{figure}

This polaron interpretation of the THz-pump/IR-probe results is corroborated by an analysis of the sign of the pump-probe signal for the two different probe energies. After the arrival of the THz pump pulse, we expect a drop of the spectral weight associated to the defect state, accompanied by an increase in the mid-infrared, large-polaron band. A schematic representation of such scenario is shown in Fig. \ref{sigma_pp} for sample S2. In Fig. \ref{sigma_pp} the optical conductivity {\it before} pump corresponds to the one measured with FTIR and already discussed in Section \ref{FTIR}, whereas the one {\it after} pump is a fictitious conductivity constructed by arbitrarily reducing the spectral weight of the defect state components, while increasing of the same amount the mid-infrared band in order to conserve the total spectral weight. A reconstruction of the corresponding transient reflectivities is shown in the lower panel of Fig. \ref{sigma_pp}, thereby highlighting that such a redistribution of the spectral weight induces a reduction of the transient reflectivity at 780 nm, while the transient reflectivity at 1560 nm increases, consistent with the experimental results.

%The previously depicted scheme also allows to perform a quantitative estimate on the photoexcited charges. Indeed, according to Fig. \ref{sigma_pp}b, in order to achieve a $\Delta R/R\sim - 6.5\cdot10^{-3}$ at 780 nm, one needs to reduce the spectral weight of the defect state component by 3.5\% . On the other hand, this would imply  $\Delta R/R\sim1.7\cdot10^{-3}$ at 1560 nm. This is in reasonable agreement with our experimental finding  $\Delta R/R\sim4.5\cdot10^{-4}$, which was measured with a pumping THz field of 1.25 MV/cm, 4 times weaker than the one employed for the 780 nm experiment.%da rivedere

%\textcolor{blue}{
%As anticipated in the previous section, the $\Delta R/R$ signal measured at 1560 nm on sample A clearly shows the activation of a coherent phonon mode. Its oscillation frequency $f_{(\lambda=1560)}=$63 GHz frequency is too low to ascribe it to an optical mode. We therefore attribute it to an acoustic phonon, where the frequency of the oscillation is related to the probe wavelength $\lambda$, to the refractive index $n$, and to the speed of sound $v_s$ through the Debye dispersion \cite{hortensius20}:
%\begin{equation}
%f=2nv_s/\lambda
%\label{acoustic}
%\end{equation}
%From equation (\ref{acoustic}), by using the refractive index $n=3.3$, % \textcolor{red}{$n=???$ n refers to the probe wavelength but it changes between 780 and 1560...} 
%as extracted from the FTIR measurement, we establish the remarkably large speed of sound $v_s=14.9$ km/s. Interestingly, an oscillation is also found in the measurement performed with probe wavelength $\lambda=780$ nm, but with a much weaker intensity. This makes the extraction of its oscillation frequency $f_{(\lambda=780)}$ much less reliable. Nonetheless, it is safe to assert that $f_{(\lambda=780)}>f_{(\lambda=1560)}$, consistent with equation (\ref{acoustic}). As far as the magnitude of the oscillation is concerned, the amplitude differences between the two wavelengths may be related to a stronger electron-phonon coupling of the polaronic MIR mode with respect to the DS contribution. Nonetheless, the larger penetration depth of the 1560 nm probe beam may also play a role in this respect.}
As anticipated in the previous section, the $\Delta R/R$ signal measured at 1560 nm on sample S2 clearly shows the activation of a coherent phonon mode. Following the theory of a damped harmonic oscillator we observe the resonance frequency $f_0$ of the phonon red shifted by the damping constant $\gamma_{\text{D}}$ = 1/$\tau_{\text{D}}$ in the following way: $f_{\text{D}}$ = $\sqrt{f_0^2 - \gamma_{\text{D}}^2 }$ = 65 GHz. By reconstructing the exponential envelope of the oscillations (see Fig. \ref{A_1.5um}b) we extract a damping constant of $\gamma_{\text{D}}$ = 54 GHz and a resonance frequency of $f_0$ = 85 GHz. The damping can be most likely ascribed to the coupling between the phonon and the polaronic charges, due to the interaction among the ions. Indeed, the significant electron-phonon coupling in a polaronic system like TiO$_{2-x}$, has a crucial role in the transient atomic vibration of the lattice upon photo-excitation \cite{mante17}.
%In Fig. \ref{A_1.5um} b) we report the signal measured at 1560 nm on sample S2 with the relative fit. The fit accounts for a rise time and two decays, which time constants are comparable with those related to the 780 nm probed signal and depicted in Fig. \ref{polaron}. 
The inset of Fig. \ref{A_1.5um}b shows the Fourier Transform of the oscillating curve reported in the main panel (in red). The arrow marks the red shift due to the damping constant. This oscillation frequency $f_0$ = 85 GHz is too low to ascribe it to an optical mode. We therefore attribute it to an acoustic eigenmode of the film.
%as it is also reported in Ref. \cite{ph_ac}, where the authors show the damped phonon frequency of 63 GHz, which is in perfect agreement with our $f_{\text{D}}$.

The frequency of the oscillation $f_0$ is related to the longitudinal sound velocity $v_s$, and to the film thickness $d$ through \cite{levchuk20}:
\begin{equation}
f_0=v_s/2d
\label{standing}
\end{equation}
From equation (\ref{standing}) we establish $v_s=$ 6800 m/s, in agreement with Ref. \cite{ph_ac}. With this value of $v_s$ we can use  $Z=\rho\cdot v_s$ \cite{weis15}, where $\rho^{TiO_2}=3900$ kg m$^{-3}$ \cite{ding14}, to estimate the acoustic impedance of the TiO$_2$ film $Z^{TiO_2}=2.7\cdot 10^{10}$ g m$^{-2}$ s$^{-1}$. The corresponding impedance for the LaAlO$_3$ substrate \cite{elias18} is $Z^{LaAlO_3}=6.5\cdot 10^{10}$ g m$^{-2}$ s$^{-1}$, thus providing an acoustic reflection coefficient $R_{ac} = (Z^{TiO_2} -Z^{LaAlO_3})/(Z^{TiO_2} + Z^{LaAlO_3})=0.41$, large enough as to sustain the standing wave formation. Interestingly, the oscillation is almost absent in the measurement performed with probe wavelength $\lambda=780$ nm. This important difference between the results obtained at the two wavelengths may be related to the strong coupling between the acoustic phonon and the polaronic mid-infrared mode, which is absent in the case of the defect state contribution. Nonetheless, the larger penetration depth of the 1560 nm probe beam may also play a role in this respect.

%An oscillation frequency of 126 GHz can also be extracted from equation \ref{acoustic} for $\lambda=780$ nm, and turns out to be compatible with the weak oscillation observed at this wavelength. Interestingly, the coherent phonon life-time at 63 GHz is much larger than the one observed at 126 GHz. The reason can be attributed to the higher penetration depth of the 1560 nm probe beam with respect to the 780 nm, meaning that a longer wavelength beam is able to probe deeper space in which the acoustic phonon propagates.

\section{Conclusions}

We provide here the first time-resolved characterization of oxygen deficient TiO$_2$ samples with THz pumping. The comparison between THz-pump/IR-probe measurement and static infrared reflectivity results allows discriminating between deep localized vacancy states and shallow delocalized polaron states. Most interestingly, THz radiation allows converting electrons from the localized state into a metastable large polaron states, with a ns lifetime. 
While shedding light on the still controversial physics underlying V$_{\text{O}}$s state in TiO$_{2-x}$, our finding has potential implications in novel electro-optic switching devices which combine different ranges of the electromagnetic spectrum. By exploiting its resistive switching properties, as well as its sensitivity to both THz and infrared light, TiO$_{2-x}$ provides the opportunity to combine sensing and processing functions, in one single device. Inspired by biological systems, TiO$_{2-x}$ based neuromorphic devices \cite{shan21} could process information from a very broad spectral range extending well above and below the visible range, with operation speed in the GHz range.

\section{Materials and methods}
\label{mm}

TiO$_2$ films were deposited at 720 °C in 10$^{-4}$ mbar oxygen environment and at a substrate-to-target distance of 7 cm, respectively, using the first harmonic of Nd:YAG laser ($\lambda$=1064 nm) \cite{Chaluvadi21, Orgiani23}. XRD characterization of the films was performed with the x-ray Panalytical Xpert high-resolution diffractometer.

FTIR reflectivity measurements at nearly normal incidence up to 1 eV were performed with a Bruker spectrometer at the SISSI-MAT beamline of the Elettra Storage-Ring \cite{lupi07}, by employing different combinations of beam-splitters and detectors. Above this frequency, in the visible and ultraviolet spectral range up to 6.5 eV, data were acquired with a Jasco V730 spectrometer at the "Sapienza" University of Rome.

THz pump pulses are provided by the TeraFERMI superradiant THz beamline at the FERMI Free-Electron Laser facility. TeraFERMI operates by utilizing relativistic sub-picosecond electron bunches to emit THz light through Coherent Transition Radiation (CTR). 

The fiber laser employed in the THz-pump/IR-probe measurements is a Menlo C-Fiber laser optically synchronized with the THz pulses. In the set-up, the THz beam is focused on the sample by a parabolic mirror with a through hole, thus allowing for normal incidence of both pump and probe beams. The THz spot size has been measured by means of a THz camera  (Spiricon, Pyrocam IIIHR), resulting in $\sim$500 $\mu$m diameter. The camera has been used to spatially overlap the THz beam and the 780/1560 nm probe beam, which hits the samples by passing through the hole of the parabolic mirror. A beamsplitter is used to both decouple the reflected light from the impinging IR beam, as well as to extract a portion of the laser beam to be used as a reference for balanced detection. 

THz electric field time trace has been measured by electro-optic sampling technique.
By combining the value of the THz spot size diameter with the THz pulse duration, as extracted by a gaussian fit of the THz electric field intensity (squared electric field time trace), the THz electric field intensity has been retrieved \cite{adhlakha23}.
The difference between the THz electric field intensities measured at 780 and 1560 nm accounts for different electron beam conditions in the FERMI accelerator during the two experiments.

The oscillations superimposed to the THz-pump/1560 nm-probe curve are extracted as the difference signal between the measured data and a double exponential fit. In order to correct for the residual nonlinear pump-probe dynamics, we applied a low pass filter to determine the baseline and substracted it. As a result, symmetric oscillations with an exponential envelope are retrieved. The curve has been smoothed with a moving average filter without broadening the oscillation itself.

\section{Acknowledgments}

The authors gratefully acknowledge the FERMI team for their essential support throughout the experiment. A.P. acknowledges G. Perucchi for the artwork in Fig. \ref{Fig1}a.    

\section{Conflict of Interest}  

The authors declare no conflict of interest.


\begin{thebibliography}{}

\bibitem{hashimoto} K. Hashimoto, H. Irie, A. Fujishima, {\it Jpn. J. Appl. Phys.} {\bf 2005}, {\it 44}, 8269.
%TiO$_2$ photocatalysis: A historical overview and future prospects
\bibitem{fujishima} A. Fujishima, K. Honda, {\it Nature} {\bf 1972}, {\it 238}, 37.
%Electrochemical photolysis of water at a semiconductor electrode
\bibitem{baldini1} E. Baldini, L. Chiodo, A. Dominguez {\it et al.}, {\it Nat. Commun.} {\bf 2017}, {\it 8}, 13.
%Strongly bound excitons in anatase TiO$_2$ single crystals and nanoparticles.
\bibitem{baldini2} E. Baldini, T. Palmieri, E. Pomarico, G. Auböck, M. Chergui,  {\it ACS Photonics} {\bf 2018}, {\it 5}, 4, 1241.
%Clocking the Ultrafast Electron Cooling in Anatase Titanium Dioxide Nanoparticles
\bibitem{baldini3} E. Baldini, T. Palmieri, A. Dominguez, P. Ruello, A. Rubio, M. Chergui, {\it Nano Letters} {\bf 2018}, {\it 18}, (8), 5007.
%Phonon-Driven Selective Modulation of Exciton Oscillator Strengths in Anatase TiO$_2$ Nanoparticles
\bibitem{baldini4} E. Baldini {\it et al.}, {\it Sci. Adv.} {\bf 2019}, {\it 5}, eaax2937.
%Exciton control in a room temperature bulk semiconductor with coherent strain pulses
\bibitem{baldini5} E. Baldini, T. Palmieri, A. Dominguez, A. Rubio, M. Chergui, {\it Phys. Rev. Lett.} {\bf 2020}, {\it 125}, 116403.
%Giant Exciton Mott Density in Anatase TiO$_2$
\bibitem{oregan} B. O’Regan,  M. Grätzel, {\it Nature} {\bf 1991}, {\it 353}, 737.
% A low-cost, high-efficiency solar cell based on dye-sensitized colloidal TiO$_2$ films
\bibitem{orgiani20} P. Orgiani {\it et al.}, {\it Phys. Rev. Applied} {\bf 2020}, {\it 13}, 044011.
%Tuning the Optical Absorption of Anatase Thin Films Across the Visible-To-Near-Infrared Spectral Region
\bibitem{bigi20} C. Bigi {\it et al.}, {\it Phys. Rev. Materials} {\bf 2020}, {\it 4}, 025801.
%Distinct behavior of localized and delocalized carriers in anatase TiO$_2$ (001) during reaction with O$_2$
\bibitem{selloni} H. Cheng, A. Selloni, {\it Phys. Rev. B} {\bf 2009}, {\it 79}, 092101.
%Surface and subsurface oxygen vacancies in anatase TiO$_2$ and differences with rutile
\bibitem{xiao22} Y. Xiao {\it et al.}, {\it Science and Technology of Advanced Materials} {\bf 2023}, {\it 24}, 2162323.
%A review of memristor: material and structure design, device performance, applications and prospects
\bibitem{lee11} M.-J. Lee, {\it Nat. Materials} {\bf 2011}, {\it 10}, 625.
%A fast, high-endurance and scalable non-volatile memory device made from asymmetric Ta$_2$O$_{5-x}$/TaO$_{2-x}$ bilayer structure
\bibitem{kousar21} F. Kousar, {\it Chaos, Solitons and Fractals} {\bf 2021}, {\it 148}, 111024.
%First principles investigation of oxygen vacancies filaments in polymorphic Titania and their role in memristor's applications
\bibitem{szot11} K. Szot {\it et al.}, {\it Nanotechnology} {\bf 2011}, {\it 22}, 254001.
%TiO$_2$-a prototypical memristive material
\bibitem{kope11} B. Magyari-K\"ope, M. Tendulkar, S.-G. Park, H.D. Lee, Y. Nishi, {\it Nanotechnology} {\bf 2011}, {\it 22}, 254029.
%Resistive switching mechanisms in random access memory devices incorporating transition metal oxides: TiO$_2$ , NiO and Pr$_{0.7}$Ca$_{0.3}$MnO$_3$
\bibitem{emin93} D. Emin, {\it Phys. Rev. B} {\bf 1993}, {\it 48}, 13691. 
%Optical properties of large and small polarons and bipolarons
\bibitem{calvani01} P. Calvani, {\it Riv. Nuovo Cimento Soc. Ital. Fis.} {\bf 2001}, {\it 24}, 1.
%Optical Properties of Polarons
\bibitem{shan21} X. Shan {\it et al.}, {\it Adv. Science} {\bf 2021}, {\it 9}, 2104632.
%Plasmonic Optoelectronic Memristor Enabling Fully Light-Modulated Synaptic Plasticity for Neuromorphic Vision
\bibitem{Knez20} D. Knez {\it et al.}, {\it Nano Lett.} {\bf 2020}, {\it 20}, 9, 6444.
% Unveiling Oxygen Vacancy Superstructures in Reduced Anatase Thin Films
\bibitem{li23} H. Li, H. Yu, X. Sun, L. Pan, {\it APL Mater.} {\bf 2023}, {\it 11}, 080601.
%Recent advances in bioinspired vision sensor arrays based on advanced optoelectronic materials
\bibitem{dressel}M. Dressel,  G. Gr\"uner, {\it Electrodynamics of Solids: optical properties of electrons in matter}, Cambridge University Press {\bf 2002}.
\bibitem{mattioli10} G. Mattioli, P. Alippi, F. Filippone, R. Caminiti, A. A. Bonapasta, {\it J. Phys. Chem. C} {\bf 2010}, {\it 114}, 21694.
% Deep versus Shallow Behavior of Intrinsic Defects in Rutile and Anatase TiO$_2$ Polymorphs
\bibitem{TeraFERMI} P. Di Pietro {\it et al.}, {\it Synchrotron Radiation News} {\bf 2017}, {\it 30}, 4, 36.
%TeraFERMI: A Superradiant Beamline for THz Nonlinear Studies at the FERMI Free Electron Laser Facility
\bibitem{perucchi13} A. Perucchi {\it et al.}, {\it Rev. Sci. Instr.} {\bf 2013}, {\it 84}, 2.
%The TeraFERMI terahertz source at the seeded FERMI free-electron-laser facility
\bibitem{Roussel} E. Roussel {\it et al.}, {\it Opt. Express} {\bf 2023}, {\it 31}, 19, 31072.
%Single-shot terahertz time-domain spectrometer using 1550 nm probe pulses and diversity electro-optic sampling
\bibitem{jerawa10} M. Jewariya, M. Nagai, K. Tanaka, {\it Phys. Rev. Lett.} {\bf 2010}, {\it 105}, 203003.
%Ladder Climbing on the Anharmonic Intermolecular Potential in an Amino Acid Microcrystal via an Intense Monocycle Terahertz Pulse
\bibitem{kornilovitch17} P.E. Kornilovitch, {\it Phys. Rev. B} {\bf 2017}, {\it 95}, 165121.
\bibitem{bretschneider18} S.A. Bretschneider, I. Ivanov, H.I. Wang, K. Miyata, X. Zhu, M. Bonn, {\it Advanced Materials} {\bf 2018}, {\it 30}, 1707312.
%Quantifying Polaron Formation and Charge Carrier Cooling in Lead-Iodide Perovskites
\bibitem{mante17} P. A. Mante, C. Stoumpos, M. Kanatzidis {\it et al.}, {\it Nat. Commun.} {\bf 2017}, {\it 8}, 14398.
% Electron–acoustic phonon coupling in single crystal CH3NH3PbI3 perovskites revealed by coherent acoustic phonons.
\bibitem{ph_ac} E.R. Cardozo de Oliveira {\it et al.}, {\it Photoacoustics} {\bf 2023}, {\it 30}, 100472.
%Probing gigahertz coherent acoustic phonons in TiO2 mesoporous thin films
\bibitem{levchuk20} A. Levchuk, B. Wilk,  G. Vaudel,  F. Labb\'e, B. Arnaud, K. Balin, J. Szade, P. Ruello , V. Juv\'e, {\it Phys. Rev. B} {\bf 2020}, {\it 101}, 180102.
%Coherent acoustic phonons generated by ultrashort terahertz pulses in nanofilms of metals and topological insulators
\bibitem{weis15} M. Weis, K. Balin,  R. Rapacz,  A. Nowak, M. Lejman,2J. Szade, P. Ruello, {\it Phys. Rev. B} {\bf 2015}, {\it 92}, 014301.
%Ultrafast light-induced coherent optical and acoustic phonons in few quintuple layers of the topological insulator Bi$_2$Te$_3$
\bibitem{ding14} Y. Ding, B. Xiao, {\it Comp. Mat. Sci.} {\bf 2014}, {\it 82}, 202.
%Anisotropic elasticity, sound velocity and thermal conductivity of TiO$_2$ polymorphs from first principles calculations
\bibitem{elias18} B. H. Elias, B. M. Ilyas, N. S. Saadi, {\it Mat. Res. Express} {\bf 2018}, {\it 5}, 086302.
%A first principle study of the perovskite lanthanum aluminate
\bibitem{Chaluvadi21} S. K. Chaluvadi {\it et al.}, {\it J. Phys. Mater.} {\bf 2021}, {\it 4}, 3, 032001.
%Pulsed laser deposition of oxide and metallic thin films by means of Nd:YAG laser source operating at its 1st harmonics: recent approaches and advances
\bibitem{Orgiani23} P. Orgiani {\it et al.}, {\it Rev. Sci. Instrum.} {\bf 2023}, {\it 94}, 3, 033903.
%Dual pulsed laser deposition system for the growth of complex materials and heterostructures
\bibitem{lupi07} S. Lupi {\it et al.},  {\it J. Opt. Soc. Am. B} {\bf 2007}, {\it 24}, 959.
%Performance of SISSI, the infrared beamline of the ELETTRA storage ring 
\bibitem{adhlakha23} N. Adhlakha, Z. Ebrahimpour, P. Di Pietro, J. Schmidt, F. Piccirilli, D.Fausti, A. Montanaro, E. Cappelluti, S. Lupi, A. Perucchi, {\it Phys. Rev. Applied} {\bf 2023}, {\it 20}, 054039.
%Terahertz saturable absorption from relativistic high-temperature thermodynamics in black phosphorus

\end{thebibliography}
\end{document}